\documentclass[12pt]{article}
\usepackage{amsmath,amssymb,amscd,amsbsy,amsgen,amsopn,amstext,
amsxtra}

\usepackage[dvips]{graphicx}

\begin{document}

\begin{center}
{\Large{\bf Explicit formula of the separability criterion for continuous variables systems}}
\end{center}
%\vskip .1 truecm
\centerline{\bf  Kazuo Fujikawa }
\vskip .1truecm
\centerline {\it Institute of Quantum Science, College of 
Science and Technology}
\centerline {\it Nihon University, Chiyoda-ku, Tokyo 101-8308, 
Japan}
\vskip 0.5 truecm

\begin{abstract}
A very explicit analytic formula of the separability criterion of two-party Gaussian systems is given. This formula is compared to the past formulation of the separability criterion of continuous variables two-party Gaussian systems.  
\end{abstract}

%\makeatletter
%\@addtoreset{equation}{section}
%\def\theequation{\thesection.\arabic{equation}}
%\makeatother

%\bodymatter
%\large
\section{Introduction and summary}
The separability criterion of continuous variables systems is important not only for theoretical interest but also for practical applications in quantum information processing. The most basic example of continuous variables systems is the two-party Gaussian system, which may be compared to the most basic two-qubit system in the case of discrete variables models. This study of the separability criterion of two-party Gaussian systems was initiated by Duan et al.~\cite{duan} and by Simon~\cite{simon}. The analyses by these authors are however based on the rather abstract "existence proofs"
and thus not easy to understand for the average workers in the field.
Moreover, these two works are based on quite different formulations and their mutual relation is not obvious at all.

We here present an explicit analytic formula of the necessary and sufficient separability criterion of two-party Gaussian systems~\cite{fujikawa1,fujikawa2}, which should be useful to the wider audience in the field. We also clarify the difference in the above two approaches~\cite{duan, simon} explicitly.\\

 To be specific, we show:\\

\noindent (i) We start with the $4\times 4$ correlation matrix $V=(V_{\mu\nu})$ 
where
\begin{eqnarray}  
V_{\mu\nu}=\frac{1}{2}
\langle\Delta\hat{\xi}_{\mu}\Delta\hat{\xi}_{\nu}+
\Delta\hat{\xi}_{\nu}\Delta\hat{\xi}_{\mu}\rangle
=\frac{1}{2}
\langle\{\Delta\hat{\xi}_{\mu}, \Delta\hat{\xi}_{\nu}\}
\rangle
\end{eqnarray}
with $\Delta\hat{\xi}_{\mu}=\hat{\xi}_{\mu}
-\langle\hat{\xi}_{\mu}\rangle$ in term of the variables
$(\hat{\xi}_{\mu})=(\hat{q}_{1}, \hat{p}_{1}, \hat{q}_{2}, 
\hat{p}_{2})$ for the two one-dimensional systems specified by canonical 
variables $(\hat{q}_{1},\hat{p}_{1})$ and 
$(\hat{q}_{2},\hat{p}_{2})$. We generally define 
$\langle\hat{O}\rangle=
{\rm Tr}\hat{\rho} \hat{O}$ by using the density matrix 
$\hat{\rho}$.

For a given standard form of covariance matrix (i.e., second moment of correlations)
\begin{eqnarray}
V_{0}=\left(\begin{array}{cccc}
  a&0&c_{1}&0\\
  0&a&0&c_{2}\\
  c_{1}&0&b&0\\
  0&c_{2}&0&b\\            
  \end{array}\right),
\end{eqnarray}
which is obtained from the general $V$ by applying the 
$Sp(2,R)\otimes Sp(2,R)$ transformations~\cite{simon},
the explicit form of {\em separability criterion} (which is in general a necessary condition) is given by 
\begin{eqnarray}
&&\ \ a\geq 1/2,\ \ b\geq 1/2, \nonumber\\
&&0\leq |c_{1}| \leq \frac{1}{2t}\{[2ab(1+t^{2})+t]-2\sqrt{D(a,b,t)}\}^{1/2},
\end{eqnarray}
where we defined 
\begin{eqnarray}
0\leq t=|c_{2}/c_{1}|\leq 1
\end{eqnarray}
without loss of generality, and
\begin{eqnarray}
&&D(a,b,t)\equiv \sqrt{a^{2}b^{2}(1-t^{2})^{2}+t(a+bt)(at+b)}.
\end{eqnarray}

\noindent (ii) For the covariance matrix obtained from the standard form $V_{0}$
by a squeezing $S^{-1}\in Sp(2,R)\otimes Sp(2,R)$ parameterized by $r_{1}$ and $r_{2}$,   
\begin{eqnarray}
V&=&S^{-1}V_{0}(S^{-1})^{T}\nonumber\\
&=&\left(\begin{array}{cccc}
  ar_{1}&0&c_{1}\sqrt{r_{1}r_{2}}&0\\
  0&a/r_{1}&0&c_{2}/\sqrt{r_{1}r_{2}}\\
  c_{1}\sqrt{r_{1}r_{2}}&0&br_{2}&0\\
  0&c_{2}/\sqrt{r_{1}r_{2}}&0&b/r_{2}\\            
  \end{array}\right), 
\end{eqnarray}
the  {\em optimal squeezing parameters} of P-representation condition 
\begin{eqnarray}
V-\frac{1}{2}I\geq0,
\end{eqnarray}
for which one can write the P-representation for the density matrix, are given by    
\begin{eqnarray}
&&r_{1}=\frac{1}{at+b}\{ab(1-t^{2})+
\sqrt{D(a,b,t)}\},\nonumber\\
&&r_{2}=\frac{1}{a+bt}\{ab(1-t^{2})+
\sqrt{D(a,b,t)}\}
\end{eqnarray}
with the same $D(a,b,t)$ in (5).
By using these parameters $r_{1}$ and $r_{2}$, one can write the the P-representation condition (7) as \begin{eqnarray}
|c_{1}|&\leq&\sqrt{(ar_{1}-1/2)(br_{2}-1/2)}/\sqrt{r_{1}r_{2}}\nonumber\\
&=&\sqrt{(a/r_{1}-1/2)(b/r_{2}-1/2)}/(t/\sqrt{r_{1}r_{2}})
\nonumber\\
&=&\frac{1}{2t}\{[2ab(1+t^{2})+t]-2\sqrt{D(a,b,t)}\}^{1/2}.
\end{eqnarray}
which agrees with (3). The P-representation defines a separable density matrix by definition, as is shown in (17) below,  
and thus the separability criterion in (3) provides the {\em necessary and sufficient separability criterion}~\cite{fujikawa1}.
\\

\section{Details of analyses}
We now discuss the details of the above analyses in connection with the past works on the separability criterion. 
\\

\noindent{\bf Simon's criterion}:\\

First of all, our formula (3) is a solution of the algebraic condition of Simon~\cite{simon}
\begin{eqnarray}
&&4(ab-c_{1}^{2})(ab-c_{2}^{2})\geq (a^{2}+b^{2})+2|c_{1}c_{2}|
-\frac{1}{4}
\end{eqnarray}
written for the standard form of covariance matrix (2). It is clear that $c_{1}=c_{2}=0$ in (2) defines a separable system, and thus we may convert (10) to a condition on $c_{1}$ and $c_{2}$. We solved the condition (10) by introducing an auxiliary parameter $t$ with $0\leq t=|c_{2}/c_{1}|\leq 1$ without loss of generality. It is however important to recognize the fact that the algebraic condition (10) allows the parameters in the range $c_{1}=c_{2}=\infty$ also, which is not allowed by our solution (3). To understand this discrepancy, one may go back to the separability criterion of Simon derived from Peres criterion~\cite{peres}
\begin{eqnarray}
\left(\begin{array}{cc}
  A&C\\
  C^{T}&B\\
            \end{array}\right)
+\frac{i}{2}\left(\begin{array}{cc}
  J&0\\
  0&\pm J\\            
  \end{array}\right)\geq 0
\end{eqnarray}
where $4\times 4$ covariance matrix $V$ is written in terms of $2\times2$ submatrices
\begin{eqnarray}
V=\left(\begin{array}{cc}
  A&C\\
  C^{T}&B\\
            \end{array}\right).
\end{eqnarray}
One can confirm that the algebraic condition (10) is given by 
\begin{eqnarray}
{\rm det}[V_{0}+\frac{i}{2}\left(\begin{array}{cc}
  J&0\\
  0&\pm J\\            
  \end{array}\right)]\geq0,
\end{eqnarray}
namely, the algebraic condition (10) does not encode the full information of the condition (11) given by Peres criterion. Note that a positive determinant does not imply a positive matrix. 
The full contents of (11) are expressed by taking the expectation value of (11) in the form $v^{\dagger}M v$ by the four-component
complex vectors $v=(d\pm ig, f\pm ih)$ with four real two-component vectors $d\sim h$ as 
\begin{eqnarray}
&&d^{T}Ad+f^{T}Bf+2d^{T}Cf+g^{T}Ag+h^{T}Bh+2g^{T}Ch
\nonumber\\
&&\geq |d^{T}Jg|+|f^{T}Jh|
\end{eqnarray}
which is $Sp(2,R)\otimes Sp(2,R)$ invariant. The condition (14) holds for any real two-component vectors $d\sim h$.

If one imposes  subsidiary conditions $g=J^{T}d$ and 
$h=\pm J^{T}f$ in (14), one obtains a {\em weaker condition} 
\begin{eqnarray}
&&d^{T}Ad+f^{T}Bf+2d^{T}Cf+d^{T}JAJ^{T}d\nonumber\\
&&+f^{T}JBJ^{T}f
\pm 2d^{T}JCJ^{T}f\nonumber\\
&&\geq (d^{T}d+f^{T}f)
\end{eqnarray}
which is no more $Sp(2,R)\otimes Sp(2,R)$ invariant. This condition (15) is easier to analyze and one obtains 
\begin{eqnarray}
\sqrt{(2a-1)(2b-1)}\geq |c_{1}|+|c_{2}|
\end{eqnarray}
with $a\geq 1/2$ and $b\geq 1/2$ for the standard form of the covariance matrix in (2). The condition (16) clearly excludes the parameter range 
$c_{1}=c_{2}=\infty$ allowed by Simon's condition (10), and we recover our condition (3).
\\

\noindent{\bf Gaussian states and P-representation}:\\

The P-representation of the density matrix 
\begin{eqnarray}
\hat{\rho}&=&\int d^{2}\alpha\int d^{2}\beta P(\alpha,\beta)
|\alpha,\beta\rangle\langle\alpha,\beta|
\end{eqnarray}
is defined in terms of coherent states and manifestly 
{\em separable}, $|\alpha,\beta\rangle=|\alpha\rangle|\beta\rangle$, and characterized by a $4\times 4$ matrix $P$ in terms of covariance matrix $V$
\begin{eqnarray}
P(\alpha,\beta)&=&\frac{\sqrt{det P}}{4\pi^{2}}
\exp\{-\frac{1}{2}(\alpha_{1},\alpha_{2},\beta_{1},\beta_{2})P(\alpha_{1},\alpha_{2},\beta_{1},\beta_{2})^{T}\}
\end{eqnarray}
where 
\begin{eqnarray}
P^{-1}=V-\frac{1}{2}I\geq 0,
\end{eqnarray}
if the P-representation exists.

The P-representation condition $V-\frac{1}{2}I\geq0$ implies in our notation in (14)
\begin{eqnarray}
d^{T}Ad+f^{T}Bf+2d^{T}Cf\geq
\frac{1}{2}(d^{T}d+f^{T}f)
\end{eqnarray}
and adding the expression with $d$ and $f$ replaced by 
$g$ and $h$ in (20), respectively, we recover the separability condition (14)
\begin{eqnarray}
&&d^{T}Ad+f^{T}Bf+2d^{T}Cf+g^{T}Ag+h^{T}Bh+2g^{T}Ch\nonumber\\
&&\geq \frac{1}{2}(d^{T}d+g^{T}g)+\frac{1}{2} (f^{T}f+h^{T}h)
\nonumber\\
&&\geq |d^{T}Jg|+|f^{T}Jh|.
\end{eqnarray}
Namely, we have shown that {\em P-representation \ $\Rightarrow$ \ separability condition} as it should since the P-representation is separable.
\\

The condition $V-\frac{1}{2}I\geq0$ is not invariant under  
$S(r_{1},r_{2})\in Sp(2,R)\otimes Sp(2,R)$, and thus
we consider the general covariance matrix in (6).

Squeezing parameters $r_{1}$ and $r_{2}$, which give the {\em boundary} of the condition $V-\frac{1}{2}I\geq0$ for given $V_{0}$,
is specified by~\cite{fujikawa1}
\begin{eqnarray}
(a-\frac{1}{2r_{1}})(b-\frac{1}{2r_{2}})
&=&\frac{1}{t^{2}}
[(a-\frac{1}{2}r_{1})(b-\frac{1}{2}r_{2})]
\end{eqnarray}
and 
\begin{eqnarray}
\frac{(ar_{1}-1/2)}{(a/r_{1}-1/2)}=
\frac{(br_{2}-1/2)}{(b/r_{2}-1/2)}.
\end{eqnarray}
These two equations are explicitly solved, and we obtain the analytic formulas of {\em optimal} squeezing parameters~\cite{fujikawa1}; 
\begin{eqnarray}
&&r_{1}=\frac{1}{at+b}\{ab(1-t^{2})+
\sqrt{D(a,b,t)}\},\nonumber\\
&&r_{2}=\frac{1}{a+bt}\{ab(1-t^{2})+
\sqrt{D(a,b,t)}\}
\end{eqnarray}
and the P-representable ({\em separable Gaussian state})  condition
$V-\frac{1}{2}I\geq0$ gives 
\begin{eqnarray}
|c_{1}|&\leq&\sqrt{(ar_{1}-1/2)(br_{2}-1/2)}/\sqrt{r_{1}r_{2}}\nonumber\\
&=&\sqrt{(a/r_{1}-1/2)(b/r_{2}-1/2)}/(t/\sqrt{r_{1}r_{2}})
\nonumber\\
&=&\frac{1}{2t}\{[2ab(1+t^{2})+t]-2\sqrt{D(a,b,t)}\}^{1/2}.
\end{eqnarray}

{\em Given} any standard form of covariance matrix $V_{0}$, we can write the separable P-representation if $|c_{1}|$ satisfies the above condition (25) for any given $a\geq \frac{1}{2},\ b\geq \frac{1}{2}$ and $1\geq t\geq 0$ by using our formulas of squeezing parameters $r_{1}$ and $r_{2}$. This establishes that our criterion in (3) provides the necessary and sufficient condition of separable Gaussian states.

Note that the squeezing, which ensures the maximum domain for $|c_{1}|$ in (25), is achieved at
$2a\geq r_{1}\geq 1,  \ \ 2b\geq r_{2}\geq 1$ to be consistent with the P-representation.
\\

\noindent{\bf Duan-Giedke-Chirac-Zoller criterion}:\\

The weaker condition (15), which is no more  $Sp(2,R)\otimes Sp(2,R)$ invariant, gives rise to the condition for the matrix (6)
\begin{eqnarray}
&&\sqrt{(ar_{1}-1/2)(br_{2}-1/2)} + \sqrt{(a/r_{1}-1/2)(b/r_{2}-1/2)}\nonumber\\
&&\geq \sqrt{r_{1}r_{2}}|c_{1}|+\frac{|c_{2}|}{\sqrt{r_{1}r_{2}}}
\end{eqnarray} 
which is in fact the original form of the separability criterion of Duan-Giedke-Chirac-Zoller~\cite{duan} based on EPR-like operators. The condition (26) is based on the condition (15), which is weaker than Simon's condition (14), {\em cannot ensure the P-representation by itself}. DGCZ then supplement their weaker condition by imposing an extra condition
\begin{eqnarray}
&&\sqrt{(ar_{1}-1/2)(br_{2}-1/2)}- \sqrt{r_{1}r_{2}}|c_{1}|\nonumber\\
&&=\sqrt{(a/r_{1}-1/2)(b/r_{2}-1/2)}-|c_{2}|/\sqrt{r_{1}r_{2}}
\end{eqnarray} 
The solution of this extra constraint in the range $2a\geq r_{1}\geq 1, \ 2b\geq r_{2}\geq 1$, if found, can ensure P-representation. But, {\em no proof} of this is given in DGCZ paper (only the existence in the interval $\infty \geq r_{1}\geq 1$ is shown), and thus their original proof is incomplete in this sense. Their proof is however completed later from a different direction~\cite{fujikawa1}.

It was also later recognized that the weaker separability criterion (26) is sufficient to ensure P-representation at the boundary of the P-representation condition. Namely, if one uses our formulas for the optimal values of squeezing parameters in (24), one can confirm that the relation (25) is equivalent to the weaker separability condition (26)~\cite{fujikawa1}. In this sense, the extra condition (27) in~\cite{duan} is not required in the analysis of separability condition of two-party Gaussian systems.   
\\

\noindent{\bf Hierarchy of separability criterions}:\\

It is also shown~\cite{fujikawa2} that we can derive a condition stronger than Simon's condition in a general context of two-party systems by an analysis of uncertainty relation ant its variants. This stronger criterion however becomes equivalent to Simon's condition for the Gaussian system. Simon's condition in turn becomes equivalent to the weaker DGCZ criterion at the boundary of the P-representation. We thus have an interesting hierarchy of seprabaility criteria for the continuous variables  two party systems.

\section{Discussion and related references}

We have presented a very explicit necessary and sufficient separability condition (3) of two-party Gaussian systems.
We finally quote some references~\cite{mancini2}-~\cite{ giovannetti} which were very helpful in 
the formulation of this explicit analytic formula.


\begin{thebibliography}{99} 
\bibitem{duan}
L.M. Duan, G. Giedke, J.I. Chirac and P. Zoller, Phys. Rev. Lett.
84 (2000) 2722.
\bibitem{simon}
R. Simon, Phys. Rev. Lett. 84 (2000) 2726.
\bibitem{fujikawa1}
K. Fujikawa, Phys. Rev. A 79 (2009) 032334.
\bibitem{fujikawa2}
K. Fujikawa, Phys. Rev. A 80 (2009) 012315. 
\bibitem{peres}
A. Peres, Phys. Rev. Lett. {\bf 77} (1996) 1413.
\bibitem{mancini2}
S. Mancini and S. Severini, Electronic Notes in Theoretical Computer Science {\bf 169} (2007) 121, and references therein.
\bibitem{shchukin}
E. Shchukin and W. Vogel, Phys. Rev. Lett. {\bf 95} (2005) 230502.
\bibitem{miranowicz}
A. Miranowicz and M. Piani, Phys. Rev. Lett. {\bf 97} (2006) 
058901.
\bibitem{werner}
R.F. Werner and M.M. Wolf, Phys. Rev. Lett. {86} (2001) 3658.
\bibitem{vidal}
G. Vidal and R.F. Werner, Phys. Rev. A{\bf 65} (2002) 032314.
\bibitem{englert}
B.G. Englert and K. Wodkiewicz, Phys. Rev. A{\bf 65} (2002)
 054303.
\bibitem{giedke}
G. Giedke, B. Kraus, M. Lewenstein, J.I. Cirac, Phys. Rev.
Lett. {\bf 87} (2001) 167904. 
\bibitem{mancini}
 S. Mancini, V. Giovannetti, D. Vitali, P. Tombesi, 
Phys. Rev. Lett. {\bf 88} (2002) 120401.
\bibitem{eisert}
J. Eisert, S. Scheel, and M.B. Plenio, Phys. Rev. Lett.{\bf 89}
(2002) 137903.
\bibitem{wolf}
M.M. Wolf, J. Eisert, and M.B. Plenio, Phys. Rev. Lett.{\bf 90}
(2003) 047904.
\bibitem{raymer}
M.G. Raymer, C. Funk, B.C. Sanders, H. de Guise, Phys. Rev. 
A{\bf 67} (2003) 052104.
\bibitem{giovannetti}
V. Giovannetti, S. Mancini, D. Vitali, P. Tombesi, Phy. Rev.
A{\bf 67} (2003) 022320.
\end{thebibliography}
\end{document}